\documentclass[journal]{vgtc}            % preprint (journal style)

\preprinttext{To appear in IEEE Transactions on Visualization and Computer Graphics.}

\ieeedoi{10.1109/VDS60365.2023.00008}

\vgtccategory{Research}

\vgtcpapertype{application/design study}

\renewcommand{\manuscriptnotetxt}{}

\graphicspath{{figs/}{figures/}{pictures/}{images/}{./}} % where to search for the images

\usepackage{tabu}% only used for the table example
\usepackage{booktabs}% only used for the table example
\usepackage{lipsum}% used to generate placeholder text
\usepackage{mwe}% used to generate placeholder figures
\usepackage{mathptmx}
\usepackage{graphicx}
\usepackage{times}
\usepackage{xspace}
\usepackage[dvipsnames,svgnames]{xcolor}
\usepackage{enumitem}
\usepackage{booktabs} % for professional tables
\usepackage{listings}
\usepackage{makecell}
\usepackage{multirow}
\usepackage{wrapfig}
\usepackage{lipsum}
\usepackage{amsfonts}
\newcommand{\toolname}{HPC$^2$lusterScape\xspace}

\newcommand{\viewnameDiagnostics}{diagnostics view\xspace}

\newcommand{\viewnameMetricTimeline}{system metrics timeline\xspace}
\newcommand{\viewnameMetricCorrelation}{bivariate analysis view\xspace}

\newcommand{\diagnosticsView}{Diagnostics view\xspace}
\newcommand{\diagnosticsViewMetricOverview}{System metric overview\xspace}
\newcommand{\diagnosticsViewMetricTimeline}{System metrics timeline\xspace}
\newcommand{\diagnosticsViewCorrelation}{Bivariate analysis between system metrics\xspace}

\newcommand{\footnoteDataDrivenDocument}{\footnote{D3 JavaScript library: https://d3js.org}\xspace}
\newcommand{\footnoteReact}{\footnote{React JavaScript framework: https://reactjs.org}\xspace}
\newcommand{\footnoteDeckgl}{\footnote{Deck.gl WebGL-powered framework: https://deck.gl}\xspace}
\newcommand{\footnoteScipy}{\footnote{SciPy Python library: https://scipy.org}\xspace}
\newcommand{\footnoteVictoriaMetrics}{\footnote{Open source Time Series Database: https://victoriametrics.com}\xspace}
\newcommand{\footnoteDcgmExporter}{\footnote{NVIDIA GPU metrics exporter: https://github.com/NVIDIA/dcgm-exporter}\xspace}
\newcommand{\footnoteNaver}{\footnote{
NAVER Co., Ltd. is the largest web search engine \& portal in South Korea.}
}
\newcommand{\footnoteKubernetes}{\footnote{
Kubernetes scheduling plugin: https://kubernetes.io/docs/reference/scheduling}
}
\title{\toolname: Increasing Transparency and Efficiency of Shared High-Performance Computing Clusters for Large-scale AI Models}
\author{%
  \authororcid{Heungseok Park*}{0000-0001-9789-8547},
  Aeree Cho*,
  Hyojun Jeon,
  Hayoung Lee,
  Youngil Yang,
  Sungjae Lee,\\
  Heungsub Lee,
  and Jaegul Choo
}

\authorfooter{
  \item
  	Heungseok Park is with NAVER Labs. E-mail: heungseok.park@naverlabs.com.
  \item
  	Aeree Cho is with Georgia Tech. E-mail: aeree@gatech.edu.
  \item
        Hyojun Jeon, Hayoung Lee, Youngil Yang, Sungjae Lee, and Heungsub Lee are with NAVER Cloud Corporation. E-mail: {hyojun.jeon, hayoung.lee, youngil.yang, sung-jae.lee, heungsub.lee}@navercorp.com.
  \item
  	Jaegul Choo is with KAIST. E-mail: jchoo@kaist.ac.kr.
   *Equally-contributing authors.
}
\abstract{
    The emergence of large-scale AI models, like GPT-4, has significantly impacted academia and industry, driving the demand for high-performance computing (HPC) to accelerate workloads. To address this, we present \toolname, a visualization system that enhances the efficiency and transparency of shared HPC clusters for large-scale AI models. \toolname provides a comprehensive overview of system-level (e.g., partitions, hosts, and workload status) and application-level (e.g., identification of experiments and researchers) information, allowing HPC operators and machine learning researchers to monitor resource utilization and identify issues through customizable violation rules. The system includes diagnostic tools to investigate workload imbalances and synchronization bottlenecks in large-scale distributed deep learning experiments. Deployed in industrial-scale HPC clusters, \toolname incorporates user feedback and meets specific requirements. This paper outlines the challenges and prerequisites for efficient HPC operation, introduces the interactive visualization system, and highlights its contributions in addressing pain points and optimizing resource utilization in shared HPC clusters.
}

\teaser{
  \centering
  \includegraphics[width=\linewidth]{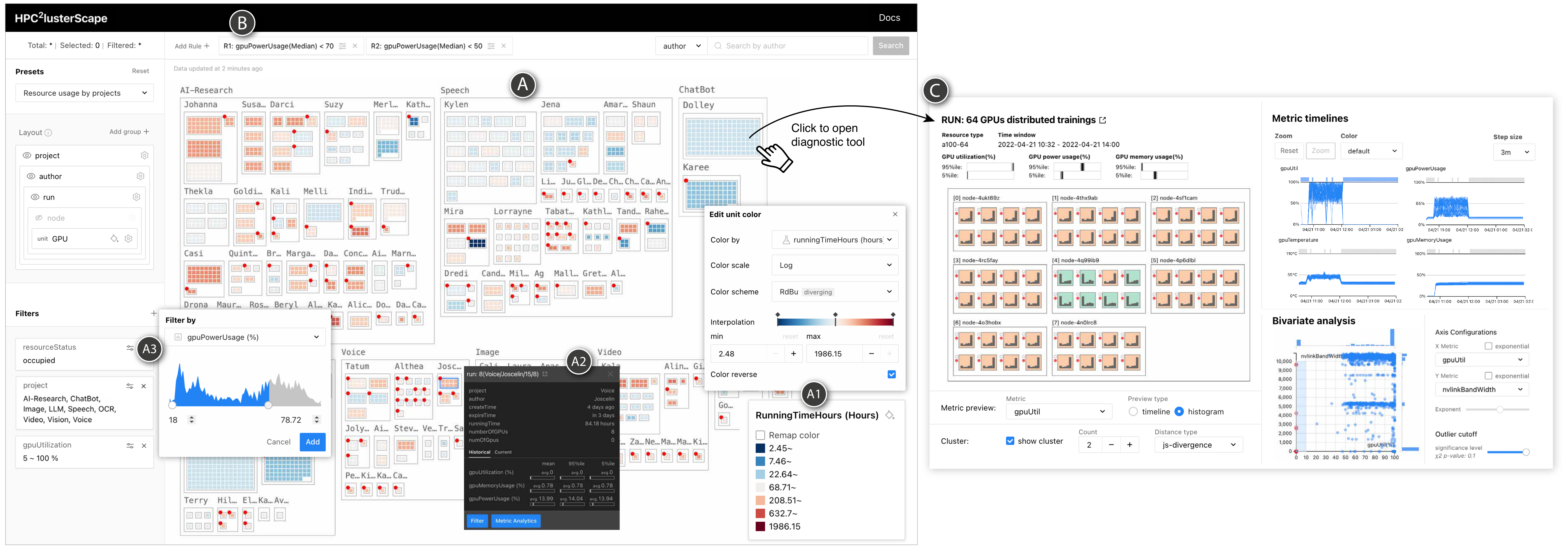}
  \vspace*{-6mm}
  \caption{%
      With \toolname, users can monitor the real-time resource status of the HPC cluster through the \textbf{(A) Cluster Overview}. GPU resources are represented as square units and can be grouped by various data attributes to provide a monitoring view that is tailored to individual needs. In this example, we have grouped GPUs by DL workloads, authors, and projects. Users can define their own conditions for the \textbf{(B) Violation Rule}, which tracks the statistics of system metrics for the workloads, determines if the rule has been violated in the past as well as at the current time, and highlights the corresponding workload. Clicking on a specific workload enables the user to enter the \textbf{(C) Diagnostics View}, which provides a node, GPU, and metric level analysis environment that allows users to drill down into that workload.
  }
  \label{fig:teaser}
  \vspace{-5pt}
}
\begin{document}

%%%%%%%%%%%%%%%%%%%%%%%%%%%%%%%%%%%%%%%%%%%%%%%%%%%%%%%%%%%%%%%%
%%%%%%%%%%%%%%%%%%%%%% START OF THE PAPER %%%%%%%%%%%%%%%%%%%%%%
%%%%%%%%%%%%%%%%%%%%%%%%%%%%%%%%%%%%%%%%%%%%%%%%%%%%%%%%%%%%%%%%

\firstsection{Introduction}
\maketitle
% \section{Introduction} %for journal use above \firstsection{..} instead
% Introduction to growing HPC and two user groups related to efficiency
The growing influence of large AI models, such as GPT, has led to a surge in demand for high-performance computing (HPC) to accelerate deep learning workloads in both industry and academia. However, effectively utilizing these high-cost resources in HPC clusters with complex hardware and network configurations requires collaborative efforts from two distinct groups: cluster operators and ML researchers. Cluster operators play a vital role in efficiently managing and operating HPC clusters. The complexity of operating such clusters lies in the need for dynamic changes in operation policies to minimize resource fragmentation and maximize training efficiency. By carefully allocating resources and addressing challenges like GPU communication bottlenecks, operators strive to optimize the utilization of HPC clusters. On the other hand, ML researchers must select resources of appropriate scale for their models and data sizes, monitor system metrics, and optimize hyperparameters and source code to make efficient use of the allocated GPUs. This is especially crucial for large-scale distributed learning, which requires numerous GPUs and extensive training times, making full resource utilization essential.

% 리소스 상태의 투명성, 유연성에 대한 중요성
However, the dynamic nature of scheduling policies, coupled with varying demands, makes wait time prediction difficult, leading to increased uncertainty in resource allocation and frustration for both operators and researchers. To address these challenges, providing transparent visibility and flexible exploration of the cluster usage and scheduling status are essential for both user groups. HPC operators need to understand how the configured policies affect resource allocation and whether they contribute to or prevent resource fragmentation. With a clear view of the scheduling state of the cluster, operators can make informed decisions when introducing policies. Similarly, by understanding the state of scheduling and potential bottlenecks, ML researchers can plan DL workload submissions and reduce unnecessary delays.
Previous studies~\cite{shilpika2019mela, guo2018valse,  dang2021hiperview, nguyen2019hiperjobviz, 9006559} primarily focused on using static views to observe cluster resources as physical units, such as nodes or racks, lacking the flexibility required for real-time exploration and decision-making. To effectively address the challenges of scheduling in shared HPC clusters, it is advantageous to provide a visualization system that offers dynamic and flexible views, empowering both HPC operators and ML researchers to monitor, analyze, and make informed decisions based on the real-time status of the cluster and scheduling policies.

% Challenges in large-scale dirstributed learning
Additionally, large-scale distributed learning inherently faces instability due to uncontrollable factors, including hardware and software failures, as well as communication overheads between different nodes~\cite{rauschmayr2022profiling}. Even minor issues can significantly impact the overall cluster efficiency given the extensive resource utilization. Collaboration with skilled operators is necessary for efficient troubleshooting and problem resolution, as ML researchers typically lack expertise in HPC cluster hardware. Previous studies have provided environments for detailed analysis of training-related metrics, system resource metrics, and GPU operation execution times~\cite{rauschmayr2022profiling, gu2017deepprof, yu2020skyline, web:pytorchprofiler, web:nsightsystems}. However, effectively identifying which aspects to analyze or where potential problems may arise when dealing with large-scale experiments involving numerous GPUs and extensive training times poses a challenge.

% Our approach
In this work, we propose a visualization system, \toolname, designed to assist two user groups in observing shared HPC clusters for efficiency enhancement. \toolname leverages unit visualization~\cite{drucker2015unifying} to represent various information of DL experiments and the shared HPC clusters, allowing both user groups can explore and monitor cluster resources from multiple perspectives along with their desired layout. \toolname provides: 1) a monitoring tool (\cref{fig:teaser}(A)) offering an integrated view of the application-level information of the ML framework (e.g., identification of experiments and researchers) and the system-level information of the HPC cluster (e.g., cluster partition, hosts, machine types, workload status), facilitating users' understanding of resource usage and scheduling status; 2) a detection tool (\cref{fig:teaser}(B)) that accumulates a diverse set of user-defined utility-violation rules with visual cues, helping users to spot potential issues of DL workloads; 3) a diagnostic tool (\cref{fig:teaser}(C)) enabling detailed analysis of problematic experiments, assisting users in identifying performance issues such as workload imbalance or synchronization bottlenecks in large-scale distributed learning. To demonstrate the utility and adaptability of the proposed system, we deploy \toolname on industrial-scale HPC clusters and conduct an evaluation through in-depth interviews with professional users, focusing on how the visualization helps two different user groups observe the HPC clusters.

The main contributions of this work are as follows: 
\begin{itemize}
    \setlength\itemsep{-0.3em}
    \item A summary of the pain points for operators and ML researchers of shared HPC clusters, factors hindering HPC efficiency, and requirements for an observational system to improve the efficiency and transparency of shared HPC clusters.
    \item An interactive observation system capable of flexibly representing both system-level and application-level information from HPC clusters and ML frameworks respectively.
    \item Detecting and diagnostic tools for identifying potential problems in large-scale distributed deep learning training.
    \item A system deployed and utilized on industrial-scale HPC clusters, demonstrating its adaptability and real-world applicability.
\end{itemize}

\section{Related Work}
\subsection{Visualization for monitoring HPC clusters}
There are numerous tools for monitoring HPC systems, such as Nagios~\cite{barth2008nagios} and Ganglia~\cite{massie2004ganglia}. However, these tools are primarily used to monitor the performance metrics of individual experiments or nodes, it's difficult to get a comprehensive view of resource usage to gain insight into scheduling status or to identify anomalous machines or experiments among a large number of resources. We reviewed studies that attempted to visualize the overall resource usage of an HPC system.

MELA~\cite{shilpika2019mela} incorporates the node layout view, representing the physical compute cards and racks of the HPC system to identify the distribution of hardware log data. VALSE~\cite{guo2018valse} employs a similar view to MELA, but with scalability through zoom interaction. The spatial view uses semantic zooming and level-of-detail rendering, ranging from compute cards to midplanes. HyperView~\cite{dang2021hiperview} is a visual analytics framework that monitors the health status of HPC systems. It represents performance metrics of nodes as radar charts and visualizes the overall distribution of nodes in a cluster as a heatmap. HyperJobViz~\cite{nguyen2019hiperjobviz} is an analytical tool designed to visualize resource allocations in data centers for jobs (i.e., workloads), users, and resource usage statistics. Waiting and running times of jobs are represented using spiral graphs, while performance metrics of nodes are displayed as radar charts. MTSAD~\cite{9006559} is a tool specifically designed for analyzing high-dimensional time series data and demonstrates its application in detecting anomalies in HPC system performance metrics. It represents performance metrics of hundreds of nodes as radar charts, and those charts that were periodically recorded are placed in a timeline graph.

Some of the aforementioned studies provide an intuitive visualization of a physical HPC cluster as a selector, along with common visualization components for in-depth analysis of selected nodes~\cite{shilpika2019mela,guo2018valse,dang2021hiperview}. Other studies visualize performance metrics on a per-node or per-job basis and allow for the comparison of multiple visualizations~\cite{nguyen2019hiperjobviz,9006559}. The former enables users to understand the physical structure of HPC systems easily, while the latter facilitates quick detection of abnormal nodes or jobs. However, since both approaches analyze HPC systems based on fixed units such as nodes or jobs, they lack the ability to integrally visualize application-level information and system-level information for HPC resource usage analysis.

\toolname provides full observability of the physical HPC cluster using unit visualization, and comprehensively presents system performance metrics, scheduling information, and application-level data using a flexible layout, color, and filter system. Furthermore, it employs user-defined violation rules to detect anomalies in system metrics of during DL workload, highlighting experiments and nodes where anomalies are detected and providing diagnostic tools for debugging.

\subsection{Visualization for analyzing computational performance of deep learning}
Numerous analytical tools have been developed to aid in profiling the performance of DL workloads, often used in conjunction with existing visualization systems.
DeepProf\cite{gu2017deepprof} is a performance analysis tool that summarizes GPU traces of TensorFlow models and generates performance analysis graphs and reports. TensorFlow (TF) Profiler\cite{web:tfprofiler} is a standalone tool integrated with TensorBoard\cite{web:tensorboard}, designed to monitor the performance of TensorFlow models by capturing operation execution times and their dependencies. PyTorch Profiler\cite{web:pytorchprofiler} provides similar functionalities for PyTorch models, allowing users to collect performance data such as input tensor shapes and stack traces. The collected data can be analyzed through the TensorBoard plugin or the Chrome Trace Viewer. On the other hand, such profiling tools have been proposed to provide deeper analysis related to the execution of CUDA kernels~\cite{web:dlprof, web:pyprof, web:nsightsystems, web:nvprof}.
These existing tools contribute to the detailed analysis of DL workload; however, they focus on low-level data analysis, leaving the entire process up to human interpretation. Furthermore, these tools lack consideration for users who may not possess expertise in hardware.

Meanwhile, research efforts have been developed to offer practitioners greater insight when analyzing DL workload performance by examining both low-level and high-level information.
% sagemaker debugger
SageMaker Debugger\cite{rauschmayr2022profiling} offers a comprehensive environment that enables practitioners to efficiently utilize resources and troubleshoot problems by discussing the most common performance issues that may occur in industry-scale DL training workflows and suggesting relevant metrics to track for each performance issue.
% SageMaker Debugger is a fully-managed service that has been deployed in the cloud-based MLOps platform and has provided a framework to automatically collect performance-related data during training.
SageMaker Debugger provided a framework to automatically collect performance-related data during training. This includes high-level metrics at the DL framework level (e.g., data loading, training, evaluation) and low-level system metrics (e.g., GPU utilization, memory usage).
Furthermore, with various visualization techniques for system metrics \cite{isaacs2014state} such as line graphs, flame graphs, and heatmaps, users can comprehensively analyze hardware issues occurring during training by cross-correlating the desired metrics.
% This environment also considered scalability so that users can efficiently analyzes synchronization issues that may occur during large-scale distributed learning.
% skyline
Skyline\cite{yu2020skyline} is an interactive in-editor tool that supports performance profiling, debugging, and visualization of DL training. It provides an environment to diagnose the performance of training by collecting and analyzing performance-related metrics such as training throughput, operation run times, and memory allocation. Furthermore, it proposes a predictive model for DL training performance based on batch size hyperparameter and offers an interface to directly modify the batch size in the in-editor environment.

The aforementioned studies contribute by suggesting metrics that impact DL training performance and providing detailed analysis at the DL framework and system resource levels. They offer valuable insights into DL training for ML researchers without extensive hardware knowledge. However, in large-scale distributed learning with numerous computations and occupied GPUs, training times are prolonged. This leads to a vast amount of data to analyze, making it difficult to visualize and analyze specific metrics like CUDA operations and memory usage individually. Moreover, system metric patterns can vary based on the learning's synchronization strategy (e.g., data/model parallelism)\cite{NIPS2012_6aca9700}, which is challenging to formalize. As a result, previous approaches like training performance prediction models have limitations.

% our approach (violation rule stacking, small multiples)
\toolname offers high-level performance information by visualizing the system resource metric distribution and pattern of each GPU used in distributed learning through small multiples. This assists users in efficiently identifying abnormal GPUs with different metric distributions. Furthermore, \toolname proposes a methodology to quickly pinpoint GPUs or nodes with varying distributions or patterns, even in cases where training involves a vast number of GPUs, by clustering them based on the time series metric data of each GPU.

\section{Identification of Factors Hindering HPC Efficiency}
\label{sec:factor_identification}

To gain a thorough understanding of shared HPC cluster operations and usage, we conducted interviews with expert users from NAVER\footnoteNaver Cloud company. Our goals were to identify efficiency obstacles, explore challenges encountered during operation and usage, and assess the potential for improved support through visualization methods.

\subsection{Background: shared HPC clusters for large-scale deep learning}
\label{sec:factor_identification_background}

Operating HPC clusters entails significant expenses in terms of acquisition, operation, maintenance, and management. Consequently, it is common for organizations or companies to share HPC clusters to optimize resource utilization. Operators of shared HPC clusters are responsible for devising scheduling systems and policies for resource allocation (i.e., DL workload submissions) and managing resources, including detecting server failures.

Resource allocation for DL workloads in AI development-oriented HPC clusters is particularly challenging. Shared HPC cluster operations often employ a mix of scheduling strategies tailored to the task nature to minimize resource waste\cite{xiao2018gandiva, narayanan2020heterogeneity, xiao2020antman, fan2021job}. Unlike the FCFS (First Come, First Serve) scheduling, allocation order in the pool is dynamically determined by the scheduling policy. In the case of large-scale distributed learning, it is even more complicated because resource allocation must take into account physical factors to minimize bottlenecks in the I/O of various sections such as CPU-GPU, GPU-GPU, and Host-Host. While virtualization and containerization technologies are designed to make it easier to manage multiple workloads on the same hardware, resource allocation for DL workloads with diverse sizes and execution times exacerbates the problem of cluster resource fragmentation.

In summary, scheduling DL workloads of varying sizes and execution times, minimizing communication bottlenecks between machines and avoiding cluster resource fragmentation is a major challenge in shared HPC operations and requires highly complex scheduling policies. These policies can change dynamically based on demand, and often require heuristics for managing waiting pools, especially in overdemand situations where spare resources are scarce.

\begin{figure*}[!h]
  \centering
  \includegraphics[width=\linewidth]{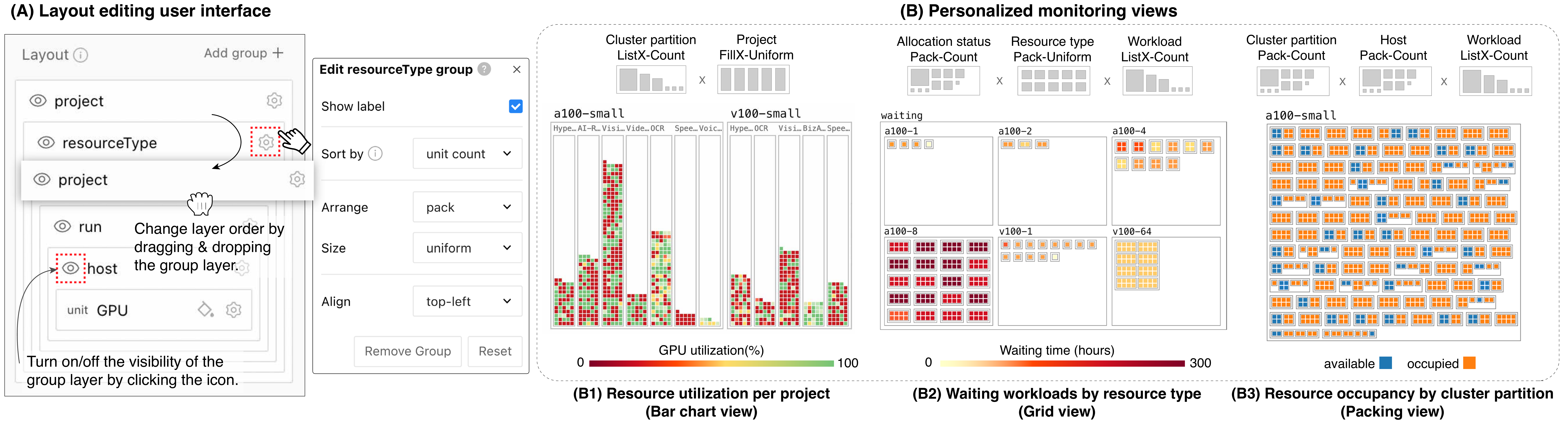}
    \vspace*{-6mm}
  \caption{%
     Demonstration of (A) layout editing user interface and (B) example monitoring views configured by various purposes.
     % With (A) Layout editing user interface, users can build (B) personalized monitoring views.
  }
  \label{fig:monitoringViews}
  \vspace{-18pt}
\end{figure*}

\subsection{Factors impeding HPC cluster efficiency}
\label{sec:factor_identification_requirements}
To understand how shared HPC clusters are commonly operated and used, as well as the obstacles to efficiency and operational challenges, we conducted semi-interviews with three HPC operators and five ML researchers from the same cluster. Participants shared their views on cluster efficiency and pain points during operation and usage. Based on these interviews, we identified key factors that negatively impact cluster efficiency and pain points that can be addressed through visualization.

\begin{enumerate}[label={\textbf{F\arabic*.}}, wide, labelwidth=!, labelindent=0pt]
    \item
    \textbf{Uncertainty and low predictability in cluster operations}
    
    During the interview process, we noted that a majority of researchers expressed frustration stemming from the uncertainty related to wait times for resource allocation. The lack of a comprehensive overview of the waiting pool and available resources presented difficulties in ascertaining the optimal resource size (i.e., the number of GPUs) or machine type (e.g., NVIDIA V100 or A100) for minimizing wait times when submitting DL workloads. Researchers suspected unfair resource allocation and abuse among teams for longer-than-expected wait times, and required transparency in resource operations.
    
    Operators acknowledged the challenges inherent in predicting and providing accurate allocation time due to intricate factors, such as workload size, machine type, and the physical location of machines. This frustration was further compounded by users' repeated inquiries regarding wait time information.
    Additionally, they highlighted the difficulty in detecting unexpected bottlenecks within resource scheduling and in understanding the impact of modifications to scheduling policies.

    \item
    \textbf{Egocentric resource monitoring}

    We noticed that researchers prioritize the success of their own experiments over improving the overall efficiency of the cluster. For instance, even when a workload's resource utilization is lower than the cluster average, researchers often spend more time adjusting hyperparameters or refining the model structure to enhance model performance, placing less importance on optimizing resource efficiency. While optimizing hyperparameters or model structure can contribute to efficiency by reducing unnecessary computations, researchers typically prioritize monitoring model performance and training stability, such as checking for memory overflow.
    
    Operators were concerned about ensuring that GPUs allocated to workloads were not being utilized (i.e., idle or inefficient workload), as it is important to maximize the overall efficiency of HPC clusters. Due to the large number of GPUs and DL workloads, they wanted to proactively identify workloads with utilization below baseline or suspected hardware failures, and they were considering how to differentiate resource allocation based on past utilization metrics per user.
    
    \item
    \textbf{Unstable large-scale training}
    
    Researchers who focus on large-scale training prioritize resource efficiency to accelerate computation during distributed workloads. However, they face challenges such as I/O bottlenecks and software/hardware failures on specific GPUs, which cause workloads to stall or synchronization to be delayed, resulting in poor training efficiency\cite{li2020taming, pumma2020alleviating, hoefler2009effect, rauschmayr2022profiling}. Despite existing strategies and studies on addressing failures in distributed learning\cite{10.1145/2783258.2783323, 7979979}, researchers often rely on operators for assistance due to limited hardware expertise, particularly regarding GPU communication topology. Therefore, there is a demand for tools that can help diagnose problems within DL workloads.

\end{enumerate}

\noindent

During the interview process, key pain points affecting shared HPC cluster efficiency were identified: 1) Operators and researchers both face challenges related to unpredictable resource allocation, due to complex policies and non-deterministic workload durations. 2) Cluster operators focus on global resource efficiency, while researchers prioritize model performance and stability, causing differing tolerances for inefficiencies. 3) In large-scale distributed learning, which consumes significant resources, both user groups seek efficiency and require detailed analysis of synchronization bottlenecks or hardware failures.

Previously proposed tools, such as monitoring of resource/scheduling status of HPC clusters\cite{massie2004ganglia, barth2008nagios, shilpika2019mela, dang2021hiperview} and analysis tools for DL training performance\cite{rauschmayr2022profiling, yu2020skyline}, can individually be seen as partially supporting the factors hindering HPC efficiency.
However, there are limitations in adequately supporting both operators and researchers, who are key user groups of HPC clusters. This drives our motivation to develop a more comprehensive design that offers transparent visibility into cluster and scheduling status, aiming to address the identified pain points.

% job? experiments? workloads?
\section{Design Goals}
\label{sec:design_goals}
In this section, we highlight and formalize the primary pain points discussed earlier in \cref{sec:factor_identification_requirements} with key design goals that \toolname aims to support. We label the three goals as G1 - G3.

\textbf{G1. Transparency of cluster \& scheduling status:} 
Our goal is to provide a transparent and comprehensive overview of resource operation status that can be tailored to the needs of both user groups. To achieve this, we aim to provide a unified visualization of system-level data and application-level data. By visualizing both levels of data, researchers can monitor the availability of GPUs and waiting pools, empowering them to plan their workload submissions effectively. Additionally, cluster operators can gain real-time insights into how the complex scheduling policy behaves. They can monitor scheduling bottlenecks and make necessary adjustments to policies based on demand. This level of transparency fosters informed decisions and helps optimize resource utilization for both researchers and operators.

\textbf{G2. Trace system metrics \& highlight abnormality:}
We aim to identify DL workloads where specific metrics deviate from predefined thresholds by leveraging GPU machine metrics during workload execution. This approach allows researchers and operators to promptly detect anomalies. Researchers can identify unstable workloads by examining significant drops or peaks in stability-related metrics, such as GPU utilization, memory usage, and temperature, both in the past and in real-time. By analyzing these metrics, researchers can pinpoint potential issues and take corrective actions. Similarly, operators can monitor metrics related to utilization and hardware failures over time to detect any abnormalities or underutilization. This helps operators promptly identify and address issues related to resource allocation or hardware failures, ensuring optimal system performance and efficiency.

\textbf{G3. Diagnosis of DL workload status at scale:}
Our goal is to provide an interface that allows users to efficiently and effectively analyze system metrics for problematic experiments. With this interface, users can identify and diagnose issues such as synchronization bottlenecks or workload/data imbalances by visualizing system metric distributions for each GPU. By comparing the distinct metric distributions among GPUs, users can detect computational or memory imbalances in large-scale distributed learning. Additionally, the interface provides analysis capabilities based on timestamps, enabling users to identify suspicious ranges and abnormal points in system metrics throughout the execution time of the workload. These features empower users to gain insights into the performance of DL workloads at scale and facilitate prompt identification and resolution of issues.

\section{HPCClusterScape Design}
\label{fig:tool_interfaces}
In this section, we describe interfaces of \toolname's, a visualization system that allows users to explore the status of shared HPC clusters and analyze the performance of large-scale AI models.

\subsection{HPC cluster overview}

\begin{figure*}[!th]
  \centering
  \includegraphics[width=\linewidth]{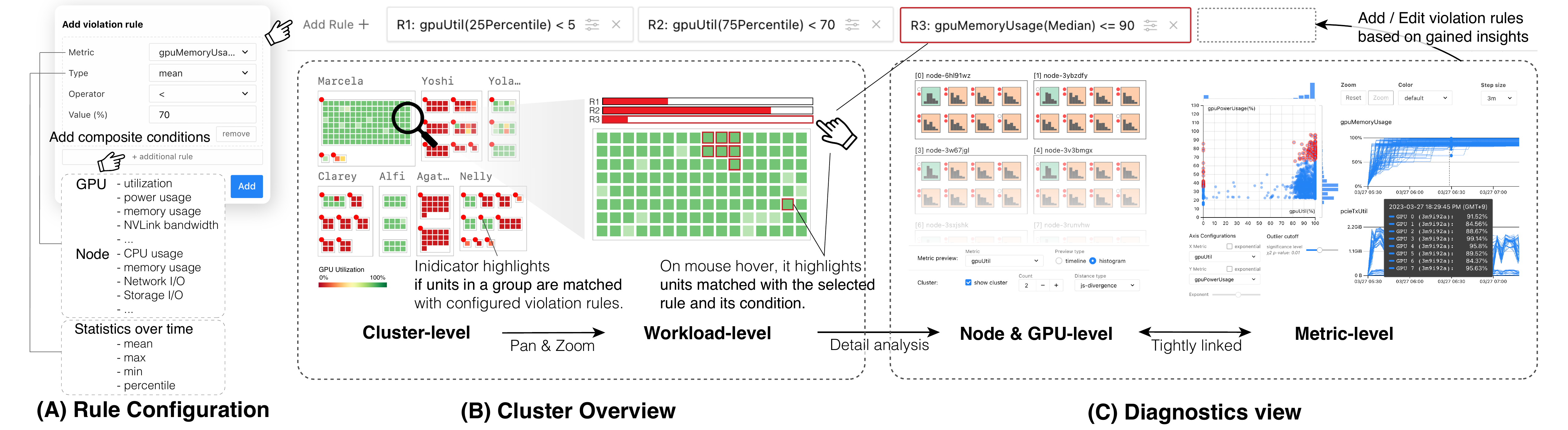}
  \vspace*{-6mm}
  \caption{%
      Representation of utility violation rules for allocated resources at various granularities. (A) Users can tailor violation rules to suit their specific DL workloads or objectives. (B) Within the cluster overview, violation rules are represented by indicators and bar plots, determined by the zoom level of the canvas. (C) In the diagnostic view for a selected DL workload, each rule is highlighted at the node and GPU levels, allowing users to examine the targeted units based on their individually configured conditions. The metric-level analysis further enhances the examination of selected units by tightly integrating with the small multiples of GPU units.
  }
  \label{fig:violation_rule_design}
  \vspace{-18pt}
\end{figure*}

\subsubsection{Multiple unit \& group layers with layout operations}

To offer multiple analytical perspectives to meet individual monitoring objectives while simultaneously presenting an overview of the cluster resource status (G1), we utilize unit visualization. Unit visualization preserves a one-to-one correspondence between rows in a data table and their representations, while still allowing for the mapping of visual attributes (e.g., color or position) onto individual units\cite{park2017atom, drucker2015unifying}. This enables the representation of both overall patterns and individual outliers. Furthermore, the visual attributes of units can be animated when transitioning from one view to another, aiding users in perceiving patterns in the data\cite{drucker2015unifying}. These benefits are particularly important given the design goal of \toolname, which aims to provide multiple views of a single data set that changes in real-time. 

The base layer of \toolname is the canvas area (\cref{fig:teaser}(A)) where units and groups are represented. A unit represents one GPU and is encoded as a square. Units are grouped based on certain data attributes and placed inside square containers, which are referred to as groups. To mitigate the typical disadvantage of visual clutter in unit visualization\cite{ellis2007taxonomy}, \toolname groups units into multiple layers without overlapping and adjusts the margin, padding, and text size differently for each layer to ensure visibility. Moreover, as shown in \cref{fig:monitoringViews}(A), the layout editing UI is designed to reveal the hierarchical structure of the group and allows for the addition and deletion of group layers via a drag-and-drop interaction, facilitating users' understanding of the layout functionality.

To construct monitoring views tailored to specific objectives, it is crucial to have the ability to flexibly arrange units and groups within each layer. For instance, a simple view grouping GPUs by projects to compare resource occupancy and a multi-layered view utilizing various hierarchical layers to monitor scheduling bottlenecks would be challenging to observe with the same layout. Users should be able to define different views by arranging units and groups within each layer. To achieve this, we redefined and implemented the layout operations proposed by Atom \cite{park2017atom}, dividing them into three categories: Pack, Fill, and List. The "Pack" operator places units and groups in a fixed aspect ratio by minimizing wasted space using a bin-packing algorithm. The "Fill" operator packs units and groups along the X or Y axis according to a fixed width or height. The "List" operator lists groups according to user-defined sorting criteria and wraps them when they are off-screen. In addition, each operation type can apply two sizes: "Uniform", which has the same size among siblings, and "Count", which depends on the size of the child elements.
Users can adjust the layout operations for each layer in the Settings pop-up UI of \cref{fig:monitoringViews}(A). \cref{fig:monitoringViews}(B1) is a layout configured to view a bar chart of how many resources projects are using per partition. Groups of projects are filled along the x-axis with the same size in a FillX-uniform layout. Projects are grouped once again by cluster partition, and these groups are listed along the X-axis in the ListX-Count layout. Along with resource occupancy by partition and project, operators can monitor which projects are utilizing resources well or poorly through color encoding.

\subsubsection{Unit color system}
Since each data attributes have different scales and distributions in their values, the min, max, and binning of the color interpolation can be adjusted to reveal meaningful information about the selected attributes. Therefore, \toolname provides color setting interface (\cref{fig:teaser}(A1)) that includes customizable color scheme and scale options such as logarithmic, quantile, and quantize. Users can also interact with a remap checkbox to adjust colors to the domain of the data filtered, which allows them to continue to leverage color for discrimination.

\subsubsection{Unit \& group interaction}
Large-scale clusters often consist of thousands of GPU and CPU devices. To enable simultaneous overview and detailed exploration, \toolname supports zooming and panning mouse interaction on the canvas area (\cref{fig:teaser}(A)). It also facilitates direct interaction with units, groups, and labels to provide on-demand details on individual unit data as well as aggregated group data. When users click the mouse, an information card UI(\cref{fig:teaser}(A2)) is displayed, which includes details about the unit or group, a summary of system metrics, and action buttons for further analysis of the workload. Additionally, the filter tool (\cref{fig:teaser}(A3)) allows users to create multiple filter conditions. For categorical data attributes, a search tool is provided for convenience.

\begin{figure*}[!t]
  \centering
  \includegraphics[width=\linewidth]{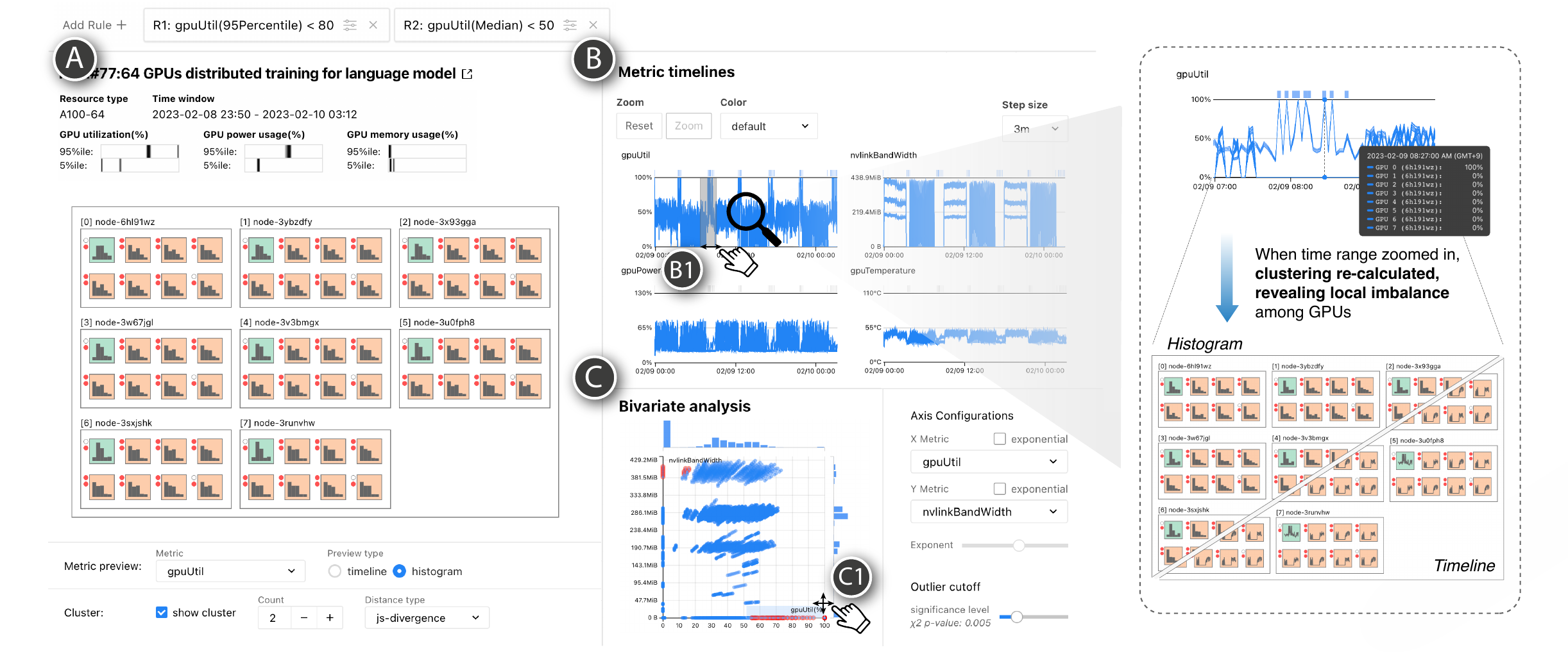}
    \vspace*{-7mm}
  \caption{%
      Diagnostics view for training status with a distributed training using 64 GPUs. (A) This view displays training status summaries, including metadata and primary metrics, while small multiples plots illustrate metric distributions for each GPU, grouped by node. The color of the small multiples plots represents the cluster's ID, determined by the distance between distributions, highlighting imbalances in the selected metric among the GPUs. Additionally, red circles beside each plot indicate activated violation rules, configured by users. (B) Line plots depict changes in system metrics over time. (C) Bivariate analysis enables users to examine correlations between selected metrics and detect outlier data points. Each view is interconnected, and when outlier points in (C) are selected using the drag interaction, the timestamp of the data points are activated on the x-axis of each line plot in (B), emphasizing the abnormal range.
  }
  \label{fig:detail_analytics}
\vspace{-17pt}
\end{figure*}

\subsection{Highlighting abnormalities based on violation rules}
\label{sec:violation_rule_interface}

By employing the unit visualization technique, the status of individual resources, as well as the resource allocation status of HPC, can be effectively represented using color encoding. This is particularly beneficial for large-scale distributed learning, as it enables the rapid identification of abnormal GPUs (see the 8 GPUs workload \texttt{Clarey} group's in \cref{fig:violation_rule_design}(B)). However, a snapshot of the current resource status does not reveal critical abnormalities that may have occurred in the past, potentially allowing erroneous experiments to go undetected and overlooked (partially addressing G2).

Deep learning training does not consistently utilize occupied resources; instead, usage varies depending on the training strategy and source code. For example, with a large batch size and extended epoch duration, GPU utilization may remain high for an extended period before briefly dropping during the validation phase, while in other cases, utilization may fluctuate frequently. Moreover, distributed learning, which necessitates communication among numerous GPUs and nodes, requires steps for data broadcasting and computation result synchronization, the pattern of system usage can be much more intricate and difficult to formalize compared to single-GPU experiments \cite{rauschmayr2022profiling}. 
Consequently, it is crucial not only to obtain an instantaneous snapshot of resource status but also to understand the extent of resource utilization from the beginning of the experiment to the present. Detecting abnormal and unstable DL workloads (G2) requires an understanding of whether there has been critical wrong usage in the past (e.g., low GPU utilization, low GPU power supply). Furthermore, since the pattern of system metrics varies across experiments, anomaly detection based solely on built-in rules defined by cluster operators does not guarantee accurate results. As such, users should be able to establish their own metrics and thresholds for tracking, based on their training strategies.

\toolname offers an interface for detecting abnormalities by tracking the history of key system metrics from the start of training to the current moment. As a real-time monitoring tool, \toolname leverages statistics of selected system metrics (e.g., min, max, mean, and percentile over time) instead of relying on computationally intensive machine learning or DNN approaches~\cite{9006559}. The UI at the top of \toolname allows users to define various rules for detecting abnormalities (\cref{fig:violation_rule_design}(A)). Each rule consists of a set of conditions, with each condition comprising four fields: a system metric, a statistic type for the metric, a relational operator, and a comparison value. Multiple rules can be added simultaneously and assessed independently. Additionally, multiple conditions can be combined to form a single rule for detecting complex cases such as where GPU utilization is high and power usage is low simultaneously.

% overview level
Due to the substantial number of GPU servers operating in large-scale HPC clusters and the numerous DL workloads allocating GPU resources, \toolname presents information granularity differently based on the level of analysis, as demonstrated in \cref{fig:violation_rule_design}. Initially, to promptly identify whether a group corresponds to a violation rule in the HPC cluster overview (\cref{fig:violation_rule_design}(B)), an indicator is encoded in the parent group if any unit object conforms to the rule, thereby highlighting the group. 
If the canvas zoom level is high, the met rules are stacked as a bar plot at the top of each group, displaying the number of violation rules and the proportion of units corresponding to each rule~\cite{yang2022the}. Hovering over each bar highlights the units associated with the rule and displays the rule's properties at the top of the screen, enabling rapid scanning for abnormal GPUs.

% detail level
% the (b) diagnostics view provides a more fine-grained level of information for more detailed analysis.
Subsequently, the diagnostics view (see \cref{sec:diagnostics}) provides a more detailed level of information for in-depth analysis (\cref{fig:violation_rule_design}(C)). It exhibits node-level and GPU-level information, arranging nodes and GPUs in a physical order using small multiples and employing indicators to illustrate the correspondence between GPU units and rules. The indicators are positioned according to the order of the violation rule item at the top of the screen. When a rule is satisfied, the indicator is mapped to its corresponding position in the complete list of rules and their order. If no rules are in effect, the indicator remains empty, indicating the absence of active rules.

Following the analysis at the node-level and GPU-level, metric-level analysis is offered to assess the status changes of each GPU and node over time. Based on the insights gained from the analysis, users can iteratively modify existing violation rule settings or add new rules.

\subsection{\diagnosticsView for training status at scale}
\label{sec:diagnostics}

\toolname offers an interface enabling users to perform in-depth analysis of the system status of resources utilized by DL experiments \cref{fig:detail_analytics}. The cluster overview, which presents the status of all resources, identifies workloads suspected of violating rules at the cluster level, while the detailed analysis view provides a node-, GPU-, and metric-level analysis environment for thorough examination of the experiment (G3), as illustrated in \cref{fig:violation_rule_design}.
Large-scale distributed workloads may involve numerous GPUs; thus, the details are provided on demand for fine-grained analysis.
Techniques such as focus$+$context or pan\&zoom can be employed to naturally link the overview and detailed analysis view~\cite{yang2022the, cockburn2009review, baudisch2002keeping}; however, due to space constraints, a detailed analysis of a large number of GPUs is presented in a separate view.
The diagnostic tool comprises three main components: (A) system metric overview, (B) system metrics timeline, and (C) bivariate analysis view.

\subsubsection{\diagnosticsViewMetricOverview}
Identifying problematic GPUs from individual performance metrics of numerous GPUs used in large-scale distributed training is challenging. Previous studies have visualized each GPU metric assigned to the experiment in a simple line graph~\cite{rauschmayr2022profiling, isaacs2014state}. However, this approach has limitations, as it requires manual inspection and interpretation of extensive datasets. To address this challenge, we have developed an interface that leverages small multiples techniques to analyze the distribution and patterns of metrics for each GPU. These small multiples are arranged based on the physical layout of the GPU servers (GPU index), global rank, and local rank (process ID) in distributed DL workloads. This interface streamlines the debugging and detection process of abnormal GPUs or nodes, facilitating more efficient analysis.

The small multiples in the interface represent time series data for system metrics collected from each GPU, and users have the flexibility to switch between different plot types to visualize metric distributions or changes over time. For example, by examining the distribution of GPU utilization, users can quickly identify workload imbalances, a critical issue in distributed learning where computation may be biased towards specific GPUs or nodes~\cite{li2020taming}. Similarly, by analyzing the distribution of GPU memory usage, users can detect data volume imbalances handled by each GPU. These imbalances are often challenging to identify at the source code level\cite{li2020taming, hoefler2009effect, pumma2020alleviating}, but become easily discernible through the representation of system metrics as histograms for each GPU.

However, When the scale of workload is immense, numerous small multiples plots can make it challenging to discern metric patterns or distribution for individual GPUs. To efficiently reveal distinct GPUs, clustering techniques are employed. If the plot type for small multiples is a histogram, clustering is computed by measuring the distance based on Jensen-Shannon divergence\cite{js-divergence} between the distributions on each GPU. If it is a timeline, clustering is computed using metrics such as Euclidean distance or correlation coefficient.
Although various clustering methods can be considered, the well-known agglomerative clustering algorithm was used to exclude randomness during clustering recalculation \cite{murtagh1983survey}. The clustering results are mapped to unique colors within each small multiples plot (\cref{fig:detail_analytics}(A)), allowing for rapid identification of distinct distributions or time-series patterns, revealing potential imbalances or abnormal patterns not evident through visualization.

When users click on a small multiple, the metric timeline and bivariate analysis views are linked (\cref{fig:detail_analytics}(B,C)), enabling a detailed analysis of the selected GPU's metric changes over time and the correlations between metrics of the GPU. The implementation of multiple selections allows users to compare and analyze metrics among various GPUs and nodes. 
Through the effective use of small multiples and analysis of time series data, it provides valuable insights into the patterns of GPU metrics, aiding users in diagnosing critical issues in large-scale distributed learning scenarios, such as workload and data imbalances.

\subsubsection{\diagnosticsViewMetricTimeline}

\toolname offers a line plot analysis environment for in-depth examination of system metric changes over time for the GPUs utilized in DL workloads. Analyzing the changes in a metric over time is an effective way to examine the performance of a machine learning model or identify problematic areas, and is a common approach across many analytics tools~\cite{rauschmayr2022profiling, web:tensorboard, isaacs2014state}.
The system metrics timeline also serves as a means to filter data for specific segments users wish to analyze. For DL training, which involves repetition of data preprocessing, training, and evaluation, we provide a zoom \& filter interface to enable detailed analysis of abnormalities within specific sections (coarse-to-fine). The filtered range is linked with the small multiples in \cref{fig:detail_analytics}(A) to identify imbalances or outliers in the zoomed window.
Moreover, large-scale distributed learning typically involves lengthy training times, making it challenging to visualize all metrics simultaneously. \toolname is implemented to sample data points and render them quickly for long training periods, and to display denser data as the time span expands.

\subsubsection{\diagnosticsViewCorrelation}
\toolname provides an interface for analyzing correlations between system metrics and identifying outlier data points that significantly deviate from the distribution of time series data across GPUs. Data points for two selected metrics are represented in a scatterplot, with outlier scores measured and highlighted for each point, as shown in \cref{fig:detail_analytics}(C). The scoring method is based on the Mahalanobis distance~\cite{de2000mahalanobis, leys2018detecting}, and users can adjust the chi-squared significance level to control the outlier cutoff according to the features or distribution of the metrics. Highlighted outliers can be selected using the mouse through the brush interaction, and the timestamp of the corresponding data point is activated on the x-axis of the system metric timeline, as shown in \cref{fig:detail_analytics}(B). This feature enables users to identify when an abnormal data point was recorded during the training process.

\section{Deployment to Industrial-scale Environment}
\label{sec:deployment}

\subsection{Deployment details}
\label{sec:deployment_details}
We deployed \toolname on the machine learning platform by NAVER Cloud called NSML~\cite{kim2018nsml, sung2017nsml}. The HPC clusters supported by this platform are mainly composed of NVIDIA A100 and V100 Tensor Core GPU servers. 
Each cluster consists of hundreds of nodes, and A100 device-type clusters are specifically designed for distributed deep learning by minimizing communication bottlenecks between GPU servers via InfiniBand and NVLink configurations.

On the other hand, each cluster is used by numerous researchers and multiple projects, and there is a need to train models of various sizes based on their specific domains. As the number of researchers surpasses the cluster size, the clusters are often in high demand. To prevent cluster fragmentation among GPU servers and maximize efficiency, operators employ various scheduling policies, such as pre-demand surveys, workload size formalization, and gang scheduling.

\subsection{Implementation details}
\label{sec:implementation details}
The server side of \toolname comprises three modules: one for gathering GPU server status and system metrics data, one for gathering application layer data (including scheduling data), and one for aggregating the collected data and delivering it to the REST API. We developed a TSDB using VictoriaMetrics\footnoteVictoriaMetrics to collect data on GPU server status and system metrics via the dcgm-exporter\footnoteDcgmExporter provided by NVIDIA. 

The client-side implementation of \toolname was developed using React\footnoteReact. The HPC cluster overview was implemented with Deck.gl\footnoteDeckgl, a WebGL framework to draw thousands of GPU units, and other modules for layout operations, filtering, and color encoding were implemented with JavaScript and D3\footnoteDataDrivenDocument. \toolname ingests Atom's visualization grammar which defines as JSON objects. Visual interfaces in the diagnostics view were implemented using D3 and SVG/Canvas elements. The clustering for imbalance analysis was created from SciPy\footnoteScipy's hierarchical clustering implementation. 

\section{Case Studies}
\label{sec:case_study}

To demonstrate how \toolname can help users understand the scheduling status of shared HPC clusters and analyze issues in distributed training, we conducted case studies in collaboration with a HPC cluster operator (denoted as P1) and n ML researcher who utilizes the ML platform where \toolname is deployed (denoted as P2).

\subsection{Adjusting scheduling policies}
\label{sec:case_study_adjusting_scheduling_policies}

On the platform the cluster is running on, users should select a type of predefined computing resource (i.e., resource type) when submitting DL workloads, which is determined by the machine type and the number of GPUs. For example, the A100-8 type uses a full host which has eight A100 GPUs. To prevent resource fragmentation, the cluster is divided into partitions to which only workloads using 8 or fewer GPUs can be assigned, and partitions to which larger workloads can be assigned.

P1 is responsible for investigating real-time resource demand to adjust resource supply per cluster partition or resolve scheduling bottlenecks. To do this, he configured the cluster overview of \toolname that shows the number of workloads waiting to be allocated for each resource type (\cref{fig:monitoringViews}(B2)).
P1 noticed a steady increase in the number of waiting workloads of the A100-8 resource type, and discovered through color encoding that these workloads generally have longer waiting times than those submitted with other resource types.

To further investigate the cause, P1 changed the layout to check the resource allocation status of the physical cluster partitions (\cref{fig:monitoringViews}(B3)). 
Looking at the partition for A100 small workloads ($\leq$ 8 GPUs), P1 observed that workloads using 1, 2, and 4 GPUs were spread across all hosts, preventing the allocation of A100-8 type workloads, which required all GPUs on a single host. To tackle this problem, P1 configured the MostAllocated\footnoteKubernetes scheduling policy, giving priority to hosts with more resources allocated. P1 continuously monitored host occupancy status in real-time, ensuring that the modified policy was working as intended, mitigating resource fragmentation.

\subsection{Detecting workload imbalance of large-scale distributed training}
\label{sec:case_study_detecting_imbalance}

ML researcher P2 is developing a Large Language Model (LLM) and train models with 64 A100 GPUs provided by the HPC cluster.
P2 organized a distributed learning experiment with each node applying data parallel and the GPUs within the nodes applying tensor parallel techniques to train the model.
P2 established two violation rules based on GPU utilization metrics to assess not only the experiment's current state but also the even and efficient utilization of GPUs during distributed learning (R1: GPU utilization 95th percentile $<$ 80\%, R2: GPU utilization median $<$ 50\%). Upon noticing his experiment was marked as a rule violation in the cluster overview, P2 proceeded to the \viewnameDiagnostics.

Within the diagnostics view's metric overview, small multiples and clustering results visualized the utilization distribution per GPU, revealing that the master GPU of all nodes belonged to the same cluster group and differed from other GPUs' utilization distributions, as shown in \cref{fig:detail_analytics}(A). Distributions for non-master GPUs were left-skewed, and violation rules set by P2 were activated for all of them, implying that a workload imbalance occurred during the training process.

To determine where the imbalance occurred, P2 employed a \viewnameMetricCorrelation, selecting data points using brush interactions that were distant in the distribution across metrics (\cref{fig:detail_analytics}(C1)). The selection resulted in outlier data points being activated on the \viewnameMetricTimeline, and P2 zoomed in on that region (\cref{fig:detail_analytics}(B1)).
% When zooming in, M1 found that the utilization values of the GPUs in that range were zero except for the master GPU on all nodes, and suspected that there was a bottleneck in the data aggregation process across the GPUs.
P2 noticed that the GPUs in the range had zero utilization values, except for the master GPU on all nodes. P2 suspected that there was a bottleneck in the data aggregation process across the GPUs. 
The updated histogram of small multiples retained the same clustering, and when the plot type was changed to timeline, P2 saw the same clustering result, confirming the imbalance occurred at that point.
P2 then did more GPU debugging and found that the workload was biased towards the master GPU during collective communication process.

\section{Use Cases}
\label{sec:user_study}
In this section, we describe various use cases of \toolname by the representative users and discuss its utility through in-depth interviews.

\subsection{Participants and study protocol}
\label{sec:user_study_interview}
To understand how \toolname supports the pain points of different user groups in shared HPC clusters described in \cref{sec:factor_identification_requirements}, We conducted interviews with HPC operators and ML researchers who operate and use shared HPC clusters where \toolname is deployed.

\par
\begingroup
\leftskip0.5em
\rightskip\leftskip
\textit{User A} is an ML researcher focused on developing speech recognition models. Their task is exploring optimized models by tuning numerous experiments with different hyperparameter sets.
\par
\endgroup

\par
\begingroup
\leftskip0.5em
\rightskip\leftskip
\textit{User B} is an ML researcher working on developing language models, with large-scale distributed learning experiments.
\par
\endgroup

\par
\begingroup
\leftskip0.5em
\rightskip\leftskip
\textit{User C} is the operator of shared HPC clusters, focusing on scheduling policy development to prevent cluster fragmentation.
\par
\endgroup

Each of the three participants had a 60-minute session. The initial 15 minutes focused on their primary tasks and typical workflows in monitoring using \toolname. The following 45 minutes involved reviewing their interaction sequences and customized layout combinations.
The interviewer aims to understand their perspectives on observing experiments and clusters and asks questions related to their intentions behind the interactions while thinking aloud.

\subsection{Key findings and insights}
We summarized the main findings and feedback from the interviews into the following criteria, by highlighting how the \toolname helps to monitor shared HPC clusters and analyze their training status, achieving our design goals.
\newline
\textbf{Understanding resource usage and scheduling bottlenecks:}
Both User A and B reported using the monitoring view to observe resource availability when submitting DL workloads. Through the visualization, User A filtered their frequently used small-scale experiment partition and monitored the running time of allocated experiments and the order of waiting experiments in real time. This enabled them to make predictions about allocation times and effectively plan their work. On the other hand, User B expressed that when they noticed their submitted experiments consistently being pushed back in the waiting pool, the visualization helped them quickly understand that experiments with the higher scheduling priority scores had been submitted, which resolved their doubts and provided clarity. These examples highlight the positive feedback on \toolname's ability to assist users in understanding scheduling situations, rather than solely waiting for resource allocations.
In addition, user B described their experience with a scheduling bottleneck, a situation where experiments in the waiting pool were not being allocated despite having available resources. They took a screenshot of the cluster overview and shared it with the operators, which allowed them to identify the bug more quickly and take appropriate action. They emphasized that visualization helped the communication process.
User C, a cluster operator, said that after introducing new scheduling policies, they were able to see side effects in real time that were not present in the simulation. For example, after deploying a scheduling policy that assigns higher priority scores to larger workloads, they observed waiting pools of each partition via the cluster overview and noticed that small workloads were constantly being pushed back and not allocated.
\newline
\textbf{Collaborative effect for workload optimization:}
User A, who was initially uninterested in utilization optimization, became motivated upon seeing their resource utilization significantly lower than the average in overall experiments. 
In addition, A, B, and C all mentioned detecting experiments with consistently low utilization records using violation rules. They informed each other about their efforts to optimize their experiments and instances where colleagues' experiment statuses seemed unstable or inefficient.
\newline
\textbf{Detecting stalled training by iteratively stacking violation rules:}
User B, whose main task is large-scale learning experiments, mentioned that stacking violation rules could catch many cases of GPU idling due to stalled training. 
Initially, they defined a simple violation rule condition to detect cases where some GPUs exhibited low utilization values during training (R1: GPU utilization 5th percentile $<$ 15\%). 
For many of the experiments detected by the rules, they could identify a common pattern in the case of GPU idling through \viewnameMetricCorrelation and \viewnameMetricTimeline, that training does not progress when utilization is high but power usage is low.
After gaining the empirical observation, they added an additional rule (R2: GPU utilization 95th percentile $>$ 90\%, R2: GPU power usage median $<$ 70\%) to capture stalled workloads characterized by consistently high utilization and low power usage.
However, they noted that identifying the root cause required the use of other debugging applications.
\newline
\textbf{Intuitively detecting imbalance in distributed training:}
User B provided positive feedback on the ability of small multiples plots to intuitively reveal workload and data imbalances that may arise during large-scale training. They mentioned that while workload imbalances are difficult to track at the code level, it was easy to identify which GPU or node exhibited computational bias through the small multiples. Additionally, they appreciated being able to grasp local imbalances for specific intervals while expanding a particular time interval.
They mainly identified the workload imbalance comparing the utilization and NVLink bandwidth usage of each GPU and assumed the data collection process between GPUs, represented by the NCCL library, as the primary cause. This assumption was confirmed by further GPU profiling in collaboration with user C, who has expertise in the hardware system. In particular, it was discovered that the load was biased towards the master node during the data collection process. To address this issue, they considered implementing techniques that could reduce the amount of data communication during training and modifying the network topology of the NCCL library.

\section{Discussion \& Future Work}

\textbf{Scalability of violation rules representation:}
Violation rules can be scaled to create composite rules by combining various conditions within each rule, but the number of configurable rules in \toolname is limited to five. This limitation arises from the restricted space between each group and group label in the cluster overview, as well as the limited space available for small multiples in the diagnostics view. We considered various indicating and piling approaches previously proposed~\cite{lekschas2020generic, lekschas2017hipiler}, but concluded that an excessive number of rules would hinder usability.
\newline
\textbf{Identifying the root cause of problems in DL workload:}
The user study showed that the \viewnameDiagnostics is effective in detecting workload imbalance and idling GPUs, but it also showed limitations in identifying the root cause of each problem. Providing analytics of kernel-level metrics required for root cause detection may be out of scope, but it gives a crucial research direction for future studies.
\newline
\textbf{Evaluation of cluster efficiency:}
Based on user studies, we found that many of pain points were addressed through transparency in shared HPC clusters, indicating that our work contributes to cluster efficiency.
However, quantitative evaluation of the effectiveness of the visualization approach remains a limitation as it is challenging to control time-dynamic variables such as the scheduler policy of the actual running cluster and the demand for resources.

\section{Conclusion}
In this paper, we presents \toolname, a visualization system designed to improve the efficiency and transparency of shared high-performance computing (HPC) clusters for AI model development. By offering an integrated monitoring tool and a set of diagnostic features, \toolname helps HPC operators and ML researchers to effectively manage resources, identify potential issues, and optimize their DL workloads. The system has been deployed and tested in an industrial-scale HPC cluster, demonstrating its adaptability and real-world applicability. By addressing the challenges and requirements of efficient HPC operation in shared environment, \toolname paves the way for more effective resource allocation and management, ultimately facilitating the advancement of large-scale AI models.

\newpage

%% if specified like this the section will be ommitted in review mode
\acknowledgments{
The authors wish to thank MLOps HPC team members, including Chanwoong Kim, Youngkwan Kim, and Adrian Kim.
}
% \vspace{-0.1cm}

\bibliographystyle{abbrv-doi-hyperref}

\bibliography{template}

\begin{thebibliography}{10}

\bibitem{web:pyprof}
A.~Agrawal and M.~Kolodziej.
\newblock {\em PyProf – PyTorch Profiling Tool
  \href{https://github.com/NVIDIA/PyProf}{github.com/NVIDIA/PyProf}}.
\newblock NVIDIA, Inc., 2021.
\newblock Last accessed 2023.

\bibitem{barth2008nagios}
W.~Barth.
\newblock {\em Nagios: System and network monitoring}.
\newblock No Starch Press, 2008.

\bibitem{baudisch2002keeping}
P.~Baudisch, N.~Good, V.~Bellotti, and P.~Schraedley.
\newblock Keeping things in context: a comparative evaluation of focus plus
  context screens, overviews, and zooming.
\newblock In {\em Proc. the ACM SIGCHI International Conference on Human
  Factors in Computing Systems (CHI)}, pp. 259--266, 2002.

\bibitem{cockburn2009review}
A.~Cockburn, A.~Karlson, and B.~B. Bederson.
\newblock A review of overview+ detail, zooming, and focus+ context interfaces.
\newblock {\em ACM Computing Surveys (CSUR)}, 41(1):1--31, 2009.

\bibitem{dang2021hiperview}
T.~Dang, N.~Nguyen, and Y.~Chen.
\newblock Hiperview: real-time monitoring of dynamic behaviors of
  high-performance computing centers.
\newblock {\em The Journal of Supercomputing}, 77:11807--11826, 2021.

\bibitem{de2000mahalanobis}
R.~De~Maesschalck, D.~Jouan-Rimbaud, and D.~L. Massart.
\newblock The mahalanobis distance.
\newblock {\em Chemometrics and intelligent laboratory systems}, 50(1):1--18,
  2000.

\bibitem{NIPS2012_6aca9700}
J.~Dean, G.~Corrado, R.~Monga, K.~Chen, M.~Devin, M.~Mao, M.~a. Ranzato,
  A.~Senior, P.~Tucker, K.~Yang, Q.~Le, and A.~Ng.
\newblock Large scale distributed deep networks.
\newblock In F.~Pereira, C.~Burges, L.~Bottou, and K.~Weinberger, eds., {\em
  Advances in Neural Information Processing Systems}, vol.~25. Curran
  Associates, Inc., 2012.

\bibitem{drucker2015unifying}
S.~Drucker and R.~Fernandez.
\newblock A unifying framework for animated and interactive unit
  visualizations.
\newblock {\em Microsoft Research}, 2015.

\bibitem{ellis2007taxonomy}
G.~Ellis and A.~Dix.
\newblock A taxonomy of clutter reduction for information visualisation.
\newblock {\em IEEE transactions on visualization and computer graphics},
  13(6):1216--1223, 2007.

\bibitem{fan2021job}
Y.~Fan.
\newblock Job scheduling in high performance computing.
\newblock {\em arXiv preprint arXiv:2109.09269}, 2021.

\bibitem{js-divergence}
B.~Fuglede and F.~Topsoe.
\newblock Jensen-shannon divergence and hilbert space embedding.
\newblock In {\em International Symposium on Information Theory, 2004. ISIT
  2004. Proceedings.}, 2004. \href{https://doi.org/10.1109/ISIT.2004.1365067}
{doi: {{%
10\hspace{.1pt}\discretionary{.}{%
}{.}\hspace{.4pt}1109\discretionary{/}{%
}{/}ISIT\hspace{.1pt}\discretionary{.}{%
}{.}\hspace{.4pt}2004\hspace{.1pt}\discretionary{.}{%
}{.}\hspace{.4pt}1365067}}}


\bibitem{web:tensorboard}
Google, Inc.
\newblock {\em TensorBoard: TensorFlow's visualization toolkit
  \href{https://tensorflow.org/tensorboard}{tensorflow.org/tensorboard}}.
\newblock Last accessed 2023.

\bibitem{web:tfprofiler}
Google, Inc.
\newblock {\em TensorFlow Profiler: A profiling and performance analysis tool
  for TensorFlow
  \href{https://tensorflow.org/guide/profiler}{tensorflow.org/guide/profiler}}.
\newblock Last accessed 2023.

\bibitem{gu2017deepprof}
J.~Gu, H.~Liu, Y.~Zhou, and X.~Wang.
\newblock Deepprof: Performance analysis for deep learning applications via
  mining gpu execution patterns.
\newblock {\em arXiv preprint arXiv:1707.03750}, 2017.

\bibitem{guo2018valse}
H.~Guo, S.~Di, R.~Gupta, T.~Peterka, and F.~Cappello.
\newblock La valse: Scalable log visualization for fault characterization in
  supercomputers.
\newblock In {\em EGPGV@ EuroVis}, pp. 91--100, 2018.

\bibitem{hoefler2009effect}
T.~Hoefler, T.~Schneider, and A.~Lumsdaine.
\newblock The effect of network noise on large-scale collective communications.
\newblock {\em Parallel processing letters}, 19(04):573--593, 2009.

\bibitem{isaacs2014state}
K.~E. Isaacs, A.~Gimenez, I.~Jusufi, T.~Gamblin, A.~Bhatele, M.~Schulz,
  B.~Hamann, and P.-T. Bremer.
\newblock State of the art of performance visualization.
\newblock {\em EuroVis (STARs)}, 2014.

\bibitem{kim2018nsml}
H.~Kim, M.~Kim, D.~Seo, J.~Kim, H.~Park, S.~Park, H.~Jo, K.~Kim, Y.~Yang,
  Y.~Kim, et~al.
\newblock {NSML}: Meet the mlaas platform with a real-world case study.
\newblock {\em arXiv preprint arXiv:1810.09957}, 2018.

\bibitem{lekschas2017hipiler}
F.~Lekschas, B.~Bach, P.~Kerpedjiev, N.~Gehlenborg, and H.~Pfister.
\newblock Hipiler: visual exploration of large genome interaction matrices with
  interactive small multiples.
\newblock {\em IEEE transactions on visualization and computer graphics},
  24(1):522--531, 2017.

\bibitem{lekschas2020generic}
F.~Lekschas, X.~Zhou, W.~Chen, N.~Gehlenborg, B.~Bach, and H.~Pfister.
\newblock A generic framework and library for exploration of small multiples
  through interactive piling.
\newblock {\em IEEE Transactions on Visualization and Computer Graphics},
  27(2):358--368, 2020.

\bibitem{leys2018detecting}
C.~Leys, O.~Klein, Y.~Dominicy, and C.~Ley.
\newblock Detecting multivariate outliers: Use a robust variant of the
  mahalanobis distance.
\newblock {\em Journal of experimental social psychology}, 74:150--156, 2018.

\bibitem{li2020taming}
S.~Li, T.~Ben-Nun, S.~D. Girolamo, D.~Alistarh, and T.~Hoefler.
\newblock Taming unbalanced training workloads in deep learning with partial
  collective operations.
\newblock In {\em Proceedings of the 25th ACM SIGPLAN Symposium on Principles
  and Practice of Parallel Programming}, pp. 45--61, 2020.

\bibitem{massie2004ganglia}
M.~L. Massie, B.~N. Chun, and D.~E. Culler.
\newblock The ganglia distributed monitoring system: design, implementation,
  and experience.
\newblock {\em Parallel Computing}, 30(7):817--840, 2004.

\bibitem{murtagh1983survey}
F.~Murtagh.
\newblock A survey of recent advances in hierarchical clustering algorithms.
\newblock {\em The computer journal}, 26(4):354--359, 1983.

\bibitem{narayanan2020heterogeneity}
D.~Narayanan, K.~Santhanam, F.~Kazhamiaka, A.~Phanishayee, and M.~Zaharia.
\newblock Heterogeneity-aware cluster scheduling policies for deep learning
  workloads.
\newblock In {\em Proceedings of the 14th USENIX Conference on Operating
  Systems Design and Implementation}, pp. 481--498, 2020.

\bibitem{nguyen2019hiperjobviz}
N.~Nguyen, T.~Dang, J.~Hass, and Y.~Chen.
\newblock Hiperjobviz: Visualizing resource allocations in high-performance
  computing center via multivariate health-status data.
\newblock In {\em 2019 IEEE/ACM Industry/University Joint International
  Workshop on Data-center Automation, Analytics, and Control (DAAC)}, pp.
  19--24. IEEE, 2019.

\bibitem{web:dlprof}
NVIDIA, Inc.
\newblock {\em DLProf
  \href{https://docs.nvidia.com/deeplearning/frameworks/dlprof-user-guide}{docs.nvidia.com/deeplearning/frameworks/dlprof-user-guide}},
  2023.
\newblock Last accessed 2023.

\bibitem{web:nsightsystems}
NVIDIA, Inc.
\newblock {\em Nsight Systems
  \href{https://developer.nvidia.com/nsight-systems}{developer.nvidia.com/nsight-systems}},
  2023.
\newblock Last accessed 2023.

\bibitem{web:nvprof}
NVIDIA, Inc.
\newblock {\em NVProf
  \href{https://docs.nvidia.com/cuda/profiler-users-guide}{docs.nvidia.com/cuda/profiler-users-guide}},
  2023.
\newblock Last accessed 2023.

\bibitem{park2017atom}
D.~Park, S.~M. Drucker, R.~Fernandez, and N.~Elmqvist.
\newblock Atom: A grammar for unit visualizations.
\newblock {\em IEEE transactions on visualization and computer graphics},
  24(12):3032--3043, 2017.

\bibitem{9006559}
V.~Pham, N.~Nguyen, J.~Li, J.~Hass, Y.~Chen, and T.~Dang.
\newblock Mtsad: Multivariate time series abnormality detection and
  visualization.
\newblock In {\em 2019 IEEE International Conference on Big Data (Big Data)},
  pp. 3267--3276, 2019.
  \href{https://doi.org/10.1109/BigData47090.2019.9006559}
{doi: {{%
10\hspace{.1pt}\discretionary{.}{%
}{.}\hspace{.4pt}1109\discretionary{/}{%
}{/}BigData47090\hspace{.1pt}\discretionary{.}{%
}{.}\hspace{.4pt}2019\hspace{.1pt}\discretionary{.}{%
}{.}\hspace{.4pt}9006559}}}


\bibitem{pumma2020alleviating}
S.~Pumma, D.~Buono, F.~Checconi, X.~Que, and W.-c. Feng.
\newblock Alleviating load imbalance in data processing for large-scale deep
  learning.
\newblock In {\em 2020 20th IEEE/ACM International Symposium on Cluster, Cloud
  and Internet Computing (CCGRID)}, pp. 262--271. IEEE, 2020.

\bibitem{web:pytorchprofiler}
The PyTorch Team.
\newblock {\em Pytorch Profiler: Profiling and optimizing pytorch
  applications}.
\newblock Last accessed 2023.

\bibitem{rauschmayr2022profiling}
N.~Rauschmayr, S.~Kama, M.~Kim, M.~Choi, and K.~Kenthapadi.
\newblock Profiling deep learning workloads at scale using amazon sagemaker.
\newblock In {\em Proc. the ACM SIGKDD International Conference on Knowledge
  Discovery and Data Mining (KDD)}, pp. 3801--3809, 2022.

\bibitem{shilpika2019mela}
F.~Shilpika, B.~Lusch, M.~Emani, V.~Vishwanath, M.~E. Papka, and K.-L. Ma.
\newblock Mela: A visual analytics tool for studying multifidelity hpc system
  logs.
\newblock In {\em Proc. the IEEE/ACM Industry/University Joint International
  Workshop on Data-center Automation, Analytics, and Control (DAAC)}, pp.
  13--18, 2019.

\bibitem{sung2017nsml}
N.~Sung, M.~Kim, H.~Jo, Y.~Yang, J.~Kim, L.~Lausen, Y.~Kim, G.~Lee, D.~Kwak,
  and J.-W. Ha.
\newblock {NSML}: A machine learning platform that enables you to focus on your
  models.
\newblock {\em arXiv preprint arXiv:1712.05902}, 2017.

\bibitem{7979979}
S.~Teerapittayanon, B.~McDanel, and H.~Kung.
\newblock Distributed deep neural networks over the cloud, the edge and end
  devices.
\newblock In {\em 2017 IEEE 37th International Conference on Distributed
  Computing Systems (ICDCS)}, pp. 328--339, 2017.
  \href{https://doi.org/10.1109/ICDCS.2017.226}
{doi: {{%
10\hspace{.1pt}\discretionary{.}{%
}{.}\hspace{.4pt}1109\discretionary{/}{%
}{/}ICDCS\hspace{.1pt}\discretionary{.}{%
}{.}\hspace{.4pt}2017\hspace{.1pt}\discretionary{.}{%
}{.}\hspace{.4pt}226}}}


\bibitem{xiao2018gandiva}
W.~Xiao, R.~Bhardwaj, R.~Ramjee, M.~Sivathanu, N.~Kwatra, Z.~Han, P.~Patel,
  X.~Peng, H.~Zhao, Q.~Zhang, et~al.
\newblock Gandiva: Introspective cluster scheduling for deep learning.
\newblock In {\em 13th USENIX Symposium on Operating Systems Design and
  Implementation (OSDI 18)}, pp. 595--610, 2018.

\bibitem{xiao2020antman}
W.~Xiao, S.~Ren, Y.~Li, Y.~Zhang, P.~Hou, Z.~Li, Y.~Feng, W.~Lin, and Y.~Jia.
\newblock Antman: Dynamic scaling on gpu clusters for deep learning.
\newblock In {\em OSDI}, pp. 533--548, 2020.

\bibitem{10.1145/2783258.2783323}
E.~P. Xing, Q.~Ho, W.~Dai, J.-K. Kim, J.~Wei, S.~Lee, X.~Zheng, P.~Xie,
  A.~Kumar, and Y.~Yu.
\newblock Petuum: A new platform for distributed machine learning on big data.
\newblock In {\em Proc. the ACM SIGKDD International Conference on Knowledge
  Discovery and Data Mining (KDD)}, KDD '15, p. 1335–1344. Association for
  Computing Machinery, New York, NY, USA, 2015.
  \href{https://doi.org/10.1145/2783258.2783323}
{doi: {{%
10\hspace{.1pt}\discretionary{.}{%
}{.}\hspace{.4pt}1145\discretionary{/}{%
}{/}2783258\hspace{.1pt}\discretionary{.}{%
}{.}\hspace{.4pt}2783323}}}


\bibitem{yang2022the}
Y.~Yang, W.~Xia, F.~Lekschas, C.~Nobre, R.~Krueger, and H.~Pfister.
\newblock The pattern is in the details: An evaluation of interaction
  techniques for locating, searching, and contextualizing details in
  multivariate matrix visualizations.
\newblock {\em Proc. the ACM SIGCHI International Conference on Human Factors
  in Computing Systems (CHI)}, 2022.

\bibitem{yu2020skyline}
G.~X. Yu, T.~Grossman, and G.~Pekhimenko.
\newblock Skyline: Interactive in-editor computational performance profiling
  for deep neural network training.
\newblock In {\em Proc. the ACM Symposium on User Interface Software and
  Technology (UIST)}, pp. 126--139, 2020.

\end{thebibliography}

\end{document}